# Measurements of the top quark mass and decay width with the D0 detector

Y. Ilchenko
*Southern Methodist University, Dallas, TX 75206, USA*

The top quark discovery in 1995 at Fermilab is one of the major proofs of the standard model (SM). Due to its unique place in SM, the top quark is an important particle for testing the theory and probing for new physics. This article presents most recent measurements of top quark properties from the D0 detector. In particular, the measurement of the top quark mass, the top antitop mass difference and the top quark decay width.

## 1. Introduction

The discovery of the top quark in 1995 [1, 2] confirmed the existence of a third generation of quarks predicted in the standard model (SM). Being the heaviest elementary particle known, the top quark appears to become an important particle in our understanding of the standard model and physics beyond it [3]. Because of its large mass the top quark has a very short lifetime, much shorter than the hadronization time. The predicted lifetime is only $3.3 \cdot 10^{-25}$s. Top quark is the only quark whose properties can be studied in isolation.

A Lorentz-invariant local Quantum Field Theory, the standard model is expected to conserve CPT [4]. Due to its unique properties, the top quark provides a perfect test of CPT invariance in the standard model. An ability to look at the quark before being hadronized allows to measure directly mass of the top quark and its antiquark. An observation of a mass difference between particle and antiparticle would indicate violation of CPT invariance.

Top quark through its radiative loop correction to the $W$ mass [5] constrains the mass of the Higgs boson. A precise measurement of the top quark mass provides useful information to the search of Higgs boson by constraining its region of possible masses. Another interesting aspect is that the top quark's Yukawa coupling to the Higgs boson is very close to unity (0.996 ± 0.006). That implies it may play a special role in the electroweak symmetry breaking mechanism.

## 2. Top quark production and decay

The top quark can be produced in two ways – in pairs via strong interactions or as a single quark via electroweak interactions. Top quark pairs at the Tevatron are produced at the collision energy of 2 TeV either through quark-antiquark annihilation (85%) or through gluon fusion (15%) (see Figure 1).

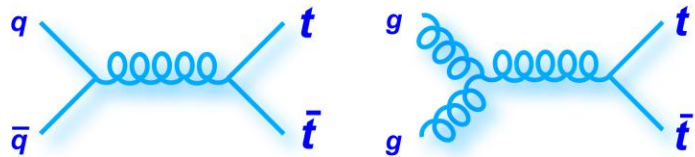

Figure 1: The top quark pair production through quark-antiquark annihilation

In SM, the top quark almost exclusively decays into a $W$ boson and $b$ quark. Subsequent decays of $W$ bosons therefore define final states of the top-antitop event. According to the decays of two $W$ bosons events can be separated by following types: all hadronic (46%), lepton+jets (30%), tau+jets(15%) and dilepton (9%).

The lepton+jets events are characterized by exactly one isolated electron or muon with large trasverse momentum, four high-$p_T$ jets and large imbalance in trasverse momentum. The main background for this channel is $W$+jets events. The lepton+jets channel provides a good combination of large statistics and distinct event signature.

The dilepton events consist of exactly two isolated leptons with large trasverse momentum, at least two high-$p_T$ jets and large imbalance in trasverse momentum. The main background for dilepton events comes from $Z$+jets events. The dilepton channel has clear event signature but suffers from low statistics.

## 3. The top quark mass

Today mass of the top quark is measured with a great precision. Currently the two most precise measurements of the mass in lepton+jets [6] and dilepton [7] channels are based on the matrix element method (ME) [8] which is discussed in detail below.

The matrix element method [8] employs all kinematic information of the event with likelihood technique. If the different physics processes that lead to the same final state do not interfere, event probability can be written as a sum of all contributing probabilities for that final state. Assuming that only a signal and dominating background (*W*+jets) contribute, the total event probability can written as

$$P_{evt} = A(x)[f \cdot P_{sig} + (1-f) \cdot P_{bkg}] \quad (1)$$

where *x* represents the measured jet and lepton energies and angles; A(x) accounts for the efficiencies and acceptance; *f* is a fraction of signal and *(1-f)* is a fraction for background in data.

Detector resolutions and the hadronization process are accounted for with the help of transfer functions, *W(x,y)*. A transfer function describes the probability of partonic final state *y* to be measured in the detector as *x*. All parameters in *W(x,y)* are derived from simulated signal events and tuned to match the resolutions observed in data.

$P_{sig}$ and $P_{bkg}$ must be integrated over all parton states contributing to the observed *x*. The probability for signal $P_{sig}$ is given by

$$P_{sig} = \frac{1}{\sigma(\alpha)} \int \sum_{flavors} dq_1 dq_2 f(q_1) f(q_2) d\sigma(y,\alpha) W(x,y) \quad (2)$$

where *α* represents the parameters to be determined (e.g. $m_t$), $f(q_i)$ are probability density functions for finding a parton of a given flavor and longitudinal momentum fraction $q_i$ in the proton or antiproton, $d\sigma(\alpha)$ is the partonic differential cross section calculated to the leading-order matrix element of the quark-antiquark annihilation process. The probability for background $P_{bkg}$ is calculated in a similar way using *W*+4 jets matrix element from the VECBOS [9] MC program.

The top quark mass is extracted from *n* events with a measured set of variables *x'*=($x_1,x_2, …, x_n$) through a likelihood function for individual event probabilities $P_{evt}$ according to

$$L(x'; m_{top}) = \prod_{i=1}^{n} P_{evt}(x_i; m_{top}, f) \quad (3)$$

The value of *f* that maximizes the likelihood is determined for every assumed $m_t$.

To calibrate the measurement results for any possible biases due to approximations used in the method, pseudoexperiment studies are performed on fully simulated Monte Carlo (MC) samples. The MC events are generated by ALPGEN program [10]. It is a matrix element based leading order (LO) generator. PYTHIA program [11] is used to describe partons to jets evolution in MC samples. GEANT3 program [12] has been applied for detector simulation and event reconstruction.

## 4. The top quark mass measurement in lepton+jets channel

The measurement of the top quark mass in lepton+jets channel is based on data corresponding to 2.6 fb$^{-1}$ of integrated luminosity. The result is combined with previous measurement in 1 fb$^{-1}$.

The matrix element technique is employed and simultaneously determines $m_t$ and *in situ* jet energy scale. The *in situ* jet energy scale accounts for a discrepancy between jets energies from *W->qq'* decay in data and Monte Carlo samples. The jets energies are therefore rescaled to the known mass of *W* boson [13]. Figure 2 shows two-dimensional likelihood $L(m_t, k_{jes})$ with fitted contours of equal probability. The likelihood is calibrated by replacing $m_t$ and $k_{jes}$ with their corrected values from the pseudoexperiment studies.

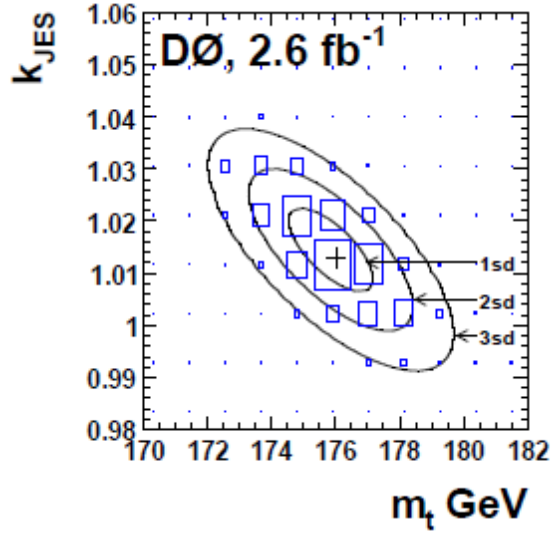

Figure 2: Likelihood $L(m_t, k_{jes})$ in lepton+jets channel with contours of equal probability

The *in situ* jet energy scale is also applied to *b* jets in the event. This results in large systematic uncertainty due to the difference in detector response between *b* and light quark jets. The response discrepancy comes from the fact that jets originating from different partons have different particle composition and kinematic characteristics. To reduce such systematic uncertainty, jet response in simulated Monte Carlo samples is corrected based on the flavor of the original parton. This allows simulated MC samples to better describe data and thereby reduce systematic uncertainty.

The combined result of the top quark mass measurement in lepton+jets events based on 3.6 fb$^{-1}$ of integrated luminosity is found to be $m_t$=174.9 ± 0.8 (stat.) ± 0.8 (jes) ± 1.0 (syst.) GeV = 174.9 ± 1.5 (stat.+syst) GeV.

## 5. The top quark mass measurement in dilepton channel

The measurement of the top quark mass in the dilepton channel is performed with 5.4 fb$^{-1}$ of integrated luminosity. This is currently the most precise measurement of $m_t$ in the dilepton channel.

As in the lepton+jets channel the matrix element technique has been used to determine $m_t$. However, unlike lepton+jets events the *in situ* jet energy scale approach cannot be applied here as there are no light-jets in the event originating from *W* decay. The flavor-dependent correction has also not been applied and therefore dominant systematic uncertainty comes from the difference in detector response between light and *b* jets.

The result is $m_t$=174.0 ± 1.8 (stat.) ± 2.4 (syst.) GeV = 176.0 ± 1.1 (stat.+syst) GeV.

## 6. Combination of the D0 top quark mass measurements

The D0 top quark mass measurement from RunI of the Tevatron are combined with results in dilepton and lepton+jets channels presented earlier here using the BLUE [14] method. We employ the same classes of uncertainties and the same program used to calculate the Tevatron average [15] of $m_t$. A summary of $m_t$ measurements used for the D0 combination and world average result are shown on Figure 3.

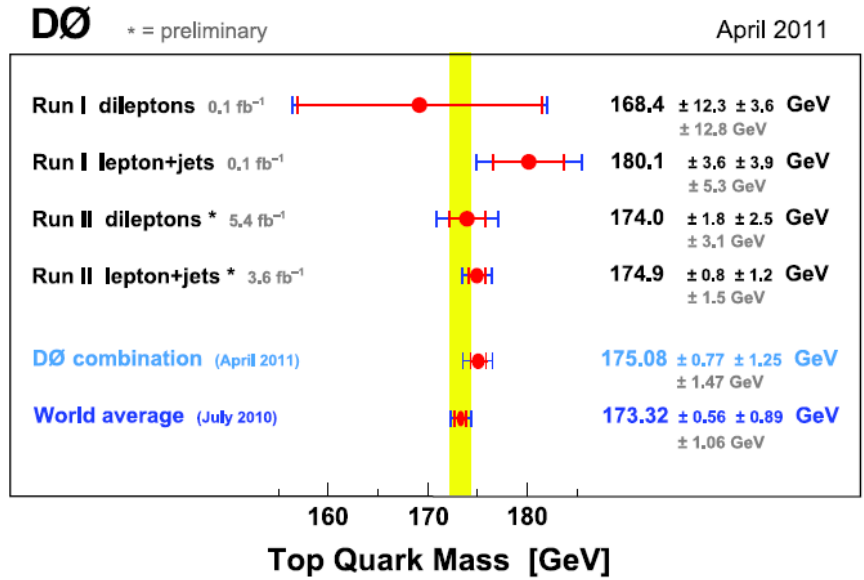

Figure 3: Top quark mass measurements at the D0, the D0 combined measurement and the world average $m_t$

## 7. The top quark mass difference

In the standard model mass of the top and antitop are required to be the same. A mass difference between those would indicate the violation of CPT invariance in the SM. The fact that the top quark decays before being hadronized makes it an ideal particle for testing CPT invariance.

We have performed a measurement of the top-antitop quark mass difference ($\Delta m$) in lepton+jets channel using same matrix element method. The measurement is done on 3.6 fb$^{-1}$ of integrated luminosity.

Simulated top-antitop quark events are required to calibrate the measurement. The PYTHIA program is modified to produce signal samples with different top-antitop masses. The signal samples are made at fourteen combinations of top and antitop quark masses with the increment of 5 GeV. The hard scattering of dominant background (*W*+jets) is simulated with ALPGEN. Partons hadronization and the shower evolution are done in PYTHIA.

The procedure of measuring the mass difference is similar to the one used to measure the top quark mass. However, in this analysis the masses of the top and antitop are measured directly instead of the top quark mass and jet energy scale. This provides simultaneous measurement of the $\Delta m$ and $m_t$.

Figure 4 shows the two-dimensional likelihood measured as a function of the top and antitop quark mass with fitted contours of equal probability. Subsequently a transformation of the likelihood to more appropriate variables $\Delta m$ and $m_{top}$ is made. Projecting transformed likelihood on the $\Delta m$ axis, the mean value $<\Delta m>$, that maximizes it, gives the best estimate of the $\Delta m$. The respective uncertainty $\delta_{\Delta m}$ on $<\Delta m>$ is calculated from the projection as well.

The combined result of the top-antitop quark mass difference in lepton+jets channel based on 3.6 fb$^{-1}$ of integrated luminosity is found to be $\Delta m = 0.8 \pm 1.8$ (stat.) $\pm 0.5$ (syst.) GeV. The result is compatible with no mass difference at the level of $\approx 1\%$ of the mass of the top quark.

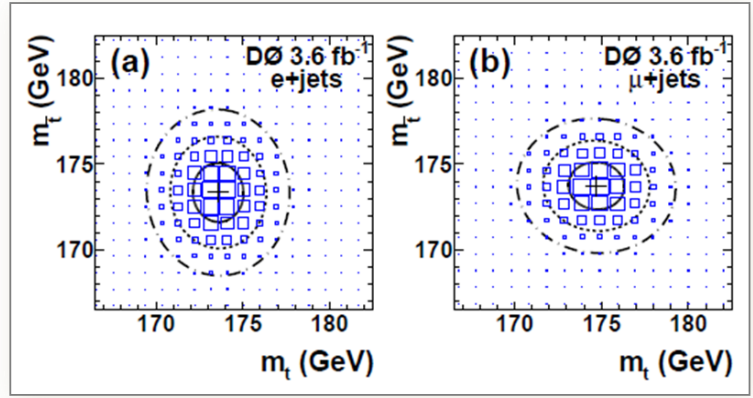

Figure 4: Two-dimensional likelihood in e+jets and µ+jets channels with contours of equal probability

## 8. The total width of the top quark

The SM predicts a very short lifetime of the top quark. The prediction results in the top quark width ($\Gamma_t$) of about 1.4 GeV at the world average top quark mass. A precise measurement of this fundamental property would allow verifying the standard model and probing for the new physics beyond it.

The direct measurement of the total width at CDF experiment set an upper bound of $\Gamma_t < 7.5$ GeV (at 95% C.L.). The width is derived from the invariant mass distribution and limited by the uncertainty on the detector resolution. The D0 experiment uses an indirect approach by separately measuring the branching fraction ratio $Br$(t→Wb) [16] and the partial width $\Gamma$(t→Wb) from single top *t*-channel cross section measurement [17]. The branching fraction ratio is found to be

$$Br(t \to Wb) = 0.962^{+0.068}_{-0.066}(stat.)^{+0.064}_{-0.052}(syst.) \qquad (4)$$

The partial width measurement is based on the assumption that the coupling in the production of the top quark is equal to the coupling in its decay. The total width becomes

$$\Gamma_t = \frac{\sigma(t-channel)\Gamma(t \to Wb)_{SM}}{Br(t \to Wb)\sigma(t-channel)_{SM}} \qquad (5)$$

The predicted SM *t*-channel cross section and $\Gamma$(t→Wb)$_{SM}$ in Equation 5 is derived from the next to leading order Monte Carlo calculations. The analysis employs the same Bayesian Neural Network discriminants as in the t-channel cross section measurement [17]. The expected and observed Bayesian probability densities for the partial width $\Gamma$(t→Wb) are shown in Figure 5. The most probable value for the partial width is defined by the peak of the probability density function and found to be

$$\Gamma(t \to Wb) = 1.92^{+0.58}_{-0.51} GeV \qquad (6)$$

The measurement of the partial width can be used to set a lower limit on the total width. Since the total width must be larger than the partial width, it must also satisfy Equation 7 at 95% C.L.

$$\Gamma_t > 1.21 GeV \qquad (7)$$

The mean value of the total width can be found by combining the partial width from Equation 7 and the branching fraction ratio from Equation 4. The total top-quark width is found to be

$$\Gamma_t = 1.99^{+0.69}_{-0.55} GeV \qquad (8)$$

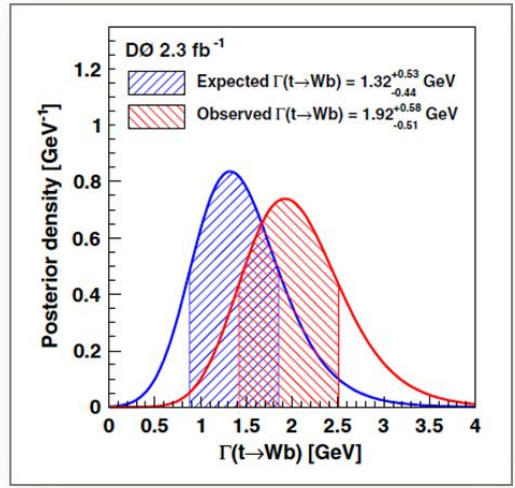

Figure 5: Probability density for the observed and expected partial width Γ(t→Wb). The hatched areas represent 1 standard deviation around the most probable value.

The total top-quark width can be expressed as the top-quark lifetime of

$$\tau = (3.3^{+1.3}_{-0.9}) \times 10^{-25} s \qquad (9)$$

The total width $\Gamma_t$ can be used to set a constraint on the coupling of a 4$^{th}$ generation *b'* quark. Assuming the *b'* quark has a high mass ($m_{b'} > m_t + m_W$), the limit on the mixing matrix element at 95% C.L. is found to be

$$|V_{tb'}| < 0.63 \qquad (10)$$

## 9. Conclusion

The most precise measurements of the top quark properties by D0 Collaboration have been presented. The high precision makes the top quark an increasingly interesting object to test the standard model and search for new physics. The combined D0 measurement of the top quark mass gives an overall uncertainty of less than 1%. The mass is found to be $m_t$=175.1 ± 0.8 (stat.) ± 1.3 (syst.) GeV = 175.1 ± 1.5 (stat.+syst) GeV. The top-anitop mass difference measurement is compatible with no mass difference at the level of 1%. The result is consistent with CPT invariance.

The total top quark width is found to be $\Gamma_t$=1.99($^{+0.69}_{-0.55}$) GeV which translates to the top-quark lifetime of τ = (1.99 $^{+0.69}_{-0.55}$)·10$^{-25}$ s. The result shows no evidence for new physics observed. In addition, the first limit on a 4$^{th}$ generation *b'* quark coupling to the top quark is set and found to be | $V_{tb'}$ | < 0.63 at 95% C.L.

## 10. Acknowledgments

I would like to thank the staffs at D0 experiment at Fermilab and collaborating institutions, and acknowledge support from DOE (USA) and Lightner-Sams Foundation.